\documentclass[10pt,conference]{IEEEtran}

\IEEEoverridecommandlockouts

\usepackage{cite}
\usepackage{amsmath,amssymb,amsfonts}
\usepackage{algorithm, algorithmic}
\usepackage{graphicx}
\usepackage{textcomp}
\usepackage{xcolor}
\usepackage{comment}
\usepackage{booktabs}
\usepackage{multirow}
\usepackage{tabularx}
\usepackage{mathtools}
\usepackage{caption}
\usepackage[normalem]{ulem}
\usepackage{subcaption}
\usepackage{soul}

\def\BibTeX{{\rm B\kern-.05em{\sc i\kern-.025em b}\kern-.08em
    T\kern-.1667em\lower.7ex\hbox{E}\kern-.125emX}}
    
\begin{document}

\title{Generalizable Multi-Task Learning for Wireless Networks Using Prompt Decision Transformers}

\author{\IEEEauthorblockN{Fatih Temiz, Shavbo Salehi, and Melike Erol-Kantarci, \textit{Fellow, IEEE}}
\IEEEauthorblockA{\textit{School of Electrical Engineering and Computer Science, University of Ottawa, Ottawa, Canada} \\
Emails: \{ftemi033, ssale038, melike.erolkantarci\}@uottawa.ca}}

\maketitle

\begin{abstract}
Future wireless networks demand rapid adaptation to highly heterogeneous environments and dynamic task configurations, necessitating a shift from conventional rule-based and optimization-driven radio resource management (RRM) toward artificial intelligence (AI)-driven RRM. AI-driven approaches can learn complex nonlinear relationships, generalize across diverse network conditions and enable real-time, scalable and autonomous decision-making. Among RRM techniques, coordinated multipoint (CoMP) transmission is pivotal for mitigating inter-cell interference and enhancing cell-edge performance, thereby improving quality of experience (QoE) in dense deployments. However, optimal multi-cell selection remains a complex combinatorial challenge as it requires jointly optimizing over many possible serving-cell combinations under dynamic traffic and channel conditions. Despite their success, conventional deep reinforcement learning (DRL) methods such as proximal policy optimization (PPO) suffer from poor sample efficiency, limited generalization, and costly retraining when state and action spaces change. To address these bottlenecks, we propose a Prompt Decision Transformer (PromptDT) based multi-task learning framework capable of learning across diverse network configurations and reformulating multi-cell selection as a sequence modeling problem. By leveraging offline trajectories and task-specific prompts, PromptDT enables scalable learning across diverse network configurations, including varying base stations and user equipment counts, and scheduler policies. Experimental results demonstrate that PromptDT improves QoE by up to 49\% in multi-task settings compared to baselines, with performance scaling positively alongside model capacity. Moreover, PromptDT generalizes effectively to unseen tasks, achieving robust few-shot adaptation to new network configurations without retraining or fine-tuning.
\end{abstract}


\section{INTRODUCTION}
The next generation of mobile networks is envisioned to host a large number of artificial intelligence (AI)-native applications. These applications can span a broad range of network functions from fault management and capacity provisioning to, most critically, radio resource management (RRM), where AI can harness the vast data generated by complex and dynamic network topologies and scenarios \cite{11370176}. 
These AI-native applications aim to overcome the inherent inefficiencies of traditional optimization methods and rigid, rule-based approaches \cite{farzanullah2025wirelessmultimodalfoundationmodel}.
In particular, reinforcement learning (RL) has long been recognized as an effective paradigm for intelligent decision-making in time-sensitive and heterogeneous wireless environments, offering significant benefits for RRM tasks such as dynamic resource allocation \cite{10699421}. 
Among these RRM techniques, coordinated multipoint (CoMP) transmission stands out as a high-impact use case for future networks, as it improves quality of experience (QoE) in dense deployments by increasing spatial diversity and mitigating inter-cell interference \cite{9427543}. However, multi-cell selection in CoMP is an NP-hard problem and remains non-trivial. 
In \cite{schneider2023deepcomp}, the authors show the benefits of a centralized RL approach, namely DeepCoMP, which is suitable for network-initiated cell-selection deployments, and demonstrate that multi-agent RL approaches support both network-initiated and mobile-originated deployments. Also, with the extensive simulations, they show that RL-based approaches significantly outperform heuristic solutions, with the centralized scheme achieving near-optimal performance by leveraging a global view of user equipments (UEs) and base stations (BSs) with minimal signaling overhead and consistently outperforming the multi-agent alternatives.

While RL offers a more flexible alternative to heuristics or integer linear programming (ILP), which typically rely on complete information of the system, conventional online RL algorithms struggle to scale in such settings. Specifically, algorithms like proximal policy optimization (PPO) suffer from poor convergence and limited generalization capabilities \cite{10839243}. For instance, simply increasing the number of BSs from three to four alters the state space, necessitating a costly retraining of the learned policy. 

To this end, large language model (LLM)-driven curriculum learning is leveraged in \cite{10682015} to improve PPO convergence, training stability, and generalization in CoMP user association under varying UE configurations. By progressively introducing more complex tasks, curriculum learning alleviates the difficulty of training in highly dynamic environments. However, despite these benefits, it still relies on iterative environment interaction and careful task sequencing. As an alternative paradigm, recent works have explored offline and sequence-modeling-based approaches that learn decision policies directly from vast historical trajectories. Among these approaches, decision transformers (DT) have recently emerged to utilize historical trajectories for sequential decision-making using advanced transformer architectures \cite{NEURIPS2021_7f489f64}. Recent studies explored the application of DT for intent-driven network management \cite{11112781}, task offloading and path planning in UAV-aided mobile edge computing (MEC) \cite{10839243}, and showed the superior performance compared to conventional RL approaches. Authors in \cite{Temiz2026EdgeLearning} adapted DT-based offline RL to a distributed setting by incorporating federated and split learning frameworks for metaverse resource allocation.
Also, the authors in \cite{ali2024explainability} examine the use of DTs for multi-cell selection in a CoMP where data is collected in a multi-agent setting. DT policy is trained and executed centrally using sequential wrapping. However, the reported performance gains were limited, primarily due to the lack of training on sufficiently diverse datasets and tasks.
 
Building upon DT, the Prompt Decision Transformer (PromptDT) is successor offline RL algorithm that leverages the sequential modeling strengths of transformer architectures and utilizes trajectory prompts for few-shot adaptation similar to LLMs \cite{NEURIPS2020_1457c0d6}. By embedding task-specific prompts, PromptDT introduces a robust and scalable multi-task learning capability, enabling the network to adapt to new configurations without the need for extensive retraining or fine-tuning\cite{xu2022prompt}. 
Recent research has begun exploring the PromptDT as a powerful alternative to traditional RL for wireless networks, though existing applications remain focused on narrow use cases. Authors in \cite{11080254} optimize resource allocation for 
MEC-enabled virtual reality streaming and show performance gains as the number of UEs varies across different environments with federated training of PromptDT. Similarly, authors in \cite{10827032}, applied for BSs energy-saving policies enabling a flexible transition when the number of BSs changes. While these works highlight the potential of PromptDT for handling environmental variations in several wireless communication scenarios, task heterogeneity is mostly restricted to a single dimension, i.e., either changes in the number of UEs or BSs. Moreover, the scalability of the PromptDT has not been explored.

To address these challenges, this paper proposes a multi-task PromptDT-based framework for complex multi-cell selection in CoMP environments. The proposed method leverages offline trajectories that are collected through diverse network configurations, encompassing variations in the number of UEs, the number of BSs, and scheduler policies. By integrating task-specific trajectory prompts with sequential decision modeling, the framework learns a flexible policy representation that enables generalization across diverse environments, supporting few-shot adaptation in a scalable manner. The key contributions of this paper are as follows:
\begin{enumerate}
    \item We propose a multi-task PromptDT framework for complex multi-cell selection in CoMP transmission where tasks vary in the numbers of UEs/BSs and scheduler policies, including resource-fair (ReF) and proportional-fair (PF), extending \cite{xu2022prompt} to multi-discrete action spaces via masked cross-entropy loss and action masking.
    \item We demonstrate robust few-shot adaptation to unseen tasks using task-specific trajectory prompts, enabling effective generalization without retraining or fine-tuning.
    \item Extensive experiments show PromptDT scales effectively with increasing task diversity and achieves up to 49\% QoE improvement as model capacity increases.
\end{enumerate}

\section{SYSTEM MODEL AND PROBLEM FORMULATION}
\begin{figure*}[h!]
    \centering
    \includegraphics[width=2\columnwidth]{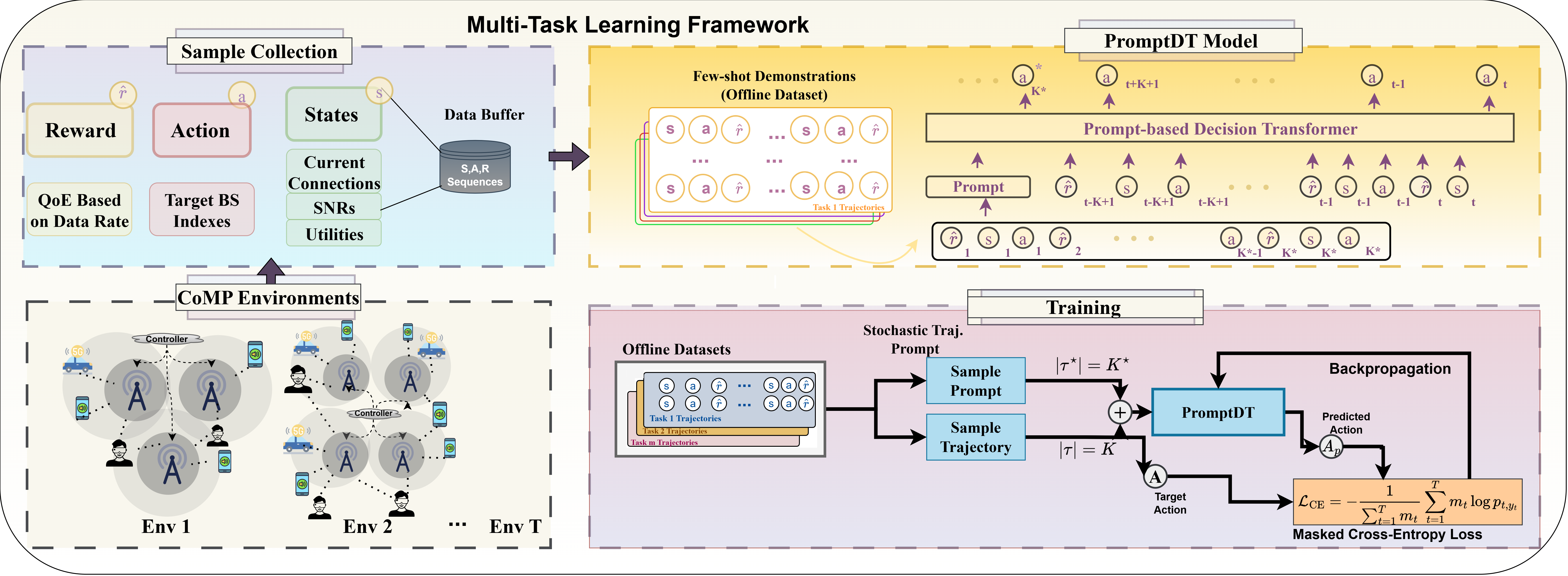}
    \caption{System Model
    }
    \label{fig:systemmodel}
\end{figure*} 
\subsection{System Model}

We consider a downlink (DL) CoMP transmission scenario in a mobile wireless network consisting of $M$ UEs and $N$ BSs, where both $M$ and $N$ can be adjusted as illustrated in Fig. \ref{fig:systemmodel}. The system is assumed to operate over discrete time steps $t = 1, 2, \dots, T$ within a bounded two-dimensional area where possible configurations are shown in Tab. \ref{tab:task_scales}. BS transmissions are assumed to be synchronized.
UE mobility follows the random waypoint model, where each UE moves with a constant velocity and periodically selects random destinations within the environment \cite{schneider2023deepcomp}. Wireless channel propagation is modeled using the Okumura-Hata path loss model, which is suitable for urban macro-cell environments. At each time step, the signal-to-noise ratio (SNR) between UE $u_j$ and BS $b_i$ is denoted by $\rho_{ij}(t)$. A UE can establish or maintain a DL connection with a BS only if the measured SNR exceeds a predefined threshold $\rho_{\min}$.

Multiple BSs are allowed to simultaneously serve a single UE, enabling CoMP transmission. BS resources i.e., physical resource blocks (PRBs) are allocated among their connected UEs using either a ReF or PF scheduling. Under ReF scheduling, each BS allocates its available PRBs equally among its connected UEs. Under PF scheduling, PRBs are distributed by balancing UEs’ instantaneous data rates with their historical throughput to achieve long-term fairness \cite{schneider2023deepcomp}.
The achievable DL data rate between BS $b_i$ and UE $u_j$ at time $t$ is denoted by $D_{ij}(t)$, and depends on the channel conditions and scheduling policy. It is obtained by aggregating the achievable DL data rates over all BSs to which the UE is connected.

To quantify the QoE of the UEs, we employ a bounded logarithmic utility function that maps the achieved DL data rate to a finite utility range. Let $D_j(t)$ denote the aggregated DL data rate received by UE $u_j$ at time $t$. The instantaneous utility of UE $u_j$ is defined as: 
\begin{equation}
U_j(t)=\min\!\left\{U_{\max},\;\max\!\left\{U_{\min},
\;
w_1 \frac{\log\!\left(w_2 + D_j(t)\right)}{\log(w_3)}
\right\}
\right\},
\end{equation}
where $w_1$, $w_2$, and $w_3$ are configurable scaling coefficients, and $U_{\min}$ and $U_{\max}$ denote the lower and upper bounds of the utility function, respectively. The clipping operation ensures that the utility remains within a bounded interval, and the rationale of this QoE formulation is to capture the diminishing returns of user-perceived satisfaction with increasing data rates.

The objective is to determine the UE-BS connectivity decisions over time to maximize the long-term average QoE across all active UEs, subject to connectivity and SNR constraints and it can be expressed as follows:
\begin{subequations}\label{eq:qoe_optimization}
\begin{equation}\label{eq:qoe_optimization_obj}
\max_{\{x_{ij}(t)\}} \;
\lim_{T\to\infty}\frac{1}{T}\sum_{t=1}^{T}\frac{1}{M(t)}
\sum_{j\in\mathcal{U}(t)} U_j(t),
\tag{\theparentequation}
\end{equation}
\begin{equation}\label{eq:qoe_optimization_bin}
x_{ij}(t)\in\{0,1\},
\quad \forall j\in\mathcal{U}(t),\ \forall i\in\mathcal{B},\ \forall t
\end{equation}
\begin{equation}\label{eq:qoe_optimization_snr}
x_{ij}(t)=0 \ \text{if}\ \rho_{ij}(t)\le \rho_{\min},
\quad \forall j\in\mathcal{U}(t),\ \forall i\in\mathcal{B},\ \forall t
\end{equation}
\begin{equation}\label{eq:qoe_optimization_conn_limit}
\sum_{i\in\mathcal{B}} x_{ij}(t) \le N,
\quad \forall j\in\mathcal{U}(t),\ \forall t
\end{equation}
\end{subequations}
where the average utility (i.e., QoE) of all active users in Eq. \ref{eq:qoe_optimization_obj} is maximized over time and Eq. \ref{eq:qoe_optimization_snr} enforces that a UE–BS connection can only be established when the SNR constraint is satisfied.
Also, Eq. \ref{eq:qoe_optimization_conn_limit} ensures that the total number of simultaneous connections for each UE does not exceed the number of BSs. 

\subsection{Problem Formulation}

Due to the stochastic evolution of user mobility and channel conditions, we model the multi-cell selection problem as a Markov Decision Process (MDP), where future system states depend only on the current state and action.
The RL agent's MDP is defined as follows:

\paragraph{State Space}
At time step $t$, the system state captures the DL connectivity, channel quality, and service utility of all active UEs. For a system with $M(t)$ active UEs and $N$ BSs, the state is defined as:
\begin{equation}
\mathbf{s}_t =
\Big[
\mathbf{x}_j(t),\;
\boldsymbol{\rho}_j(t),\;
\hat{U}_j(t)
\Big]_{j\in\mathcal{U}(t)},
\end{equation}
$\mathbf{x}_j(t)\in\{0,1\}^{N}$ denotes the binary connection vector where the $n$-th element equals $1$ if UE $u_j$ is connected to BS $b_n$, and $0$ otherwise. $\boldsymbol{\rho}_j(t)\in\mathbb{R}^{N}$ represents the normalized SNRs between UE $u_j$ and all BSs, and $\hat{U}_j(t)\in[-1,1]$ is the scaled utility of UE $u_j$.

\paragraph{Action Space}
To limit protocol overhead and signaling complexity, each UE is allowed to modify at most one connection per time step. Accordingly, the action for UE $u_j$ at time $t$ is defined as:

\begin{equation}
a_j(t)\in\{0,1,\dots,N\},
\end{equation}
where $a_j(t)=i\in\{1,\dots,N\}$ indicates that UE $u_j$ toggles its connection status with BS $b_i$, i.e., a connection is established if none exists or released otherwise, provided that the SNR constraint is satisfied. Alternatively, $a_j(t)=0$ denotes a non-operation, where all existing connections of UE $u_j$ remain unchanged. Then the joint action at time $t$ is given by $\mathbf{a}_j = \{a_j(t)\}_{j\in\mathcal{U}(t)}$.

\paragraph{Reward Function}
The reward function at the time step $t$ is defined as the average utility of all active UEs, reflecting the overall system's QoE:
\begin{equation}
r_t =
\frac{1}{M(t)}\sum_{j\in\mathcal{U}(t)} \hat{U}_j(t),
\end{equation}
This reward formulation encourages policies that maximize the long-term average QoE across users.

\section{PromptDT-based Multi-Task Learning}
The architecture of PromptDT is built upon DT that formulates RL as a sequence modeling problem where decision-making is learned through autoregressive modeling rather than an explicit value-function or policy optimization task. Contextual MDP is defined as $(\mathcal{S}, \mathcal{A}, \mathcal{P}, \mathcal{R})$ with states $s_t \in \mathcal{S}$, actions $a_t \in \mathcal{A}$, rewards $r_t = \mathcal{R}_k(s_t, a_t)$, and $\mathcal{P}(s'|s,a)$ is the transition dynamics. 
The model uses undiscounted return-to-go (RTG), i.e., the sum of future rewards, $\hat{R}_t^k = \sum_{t'=t}^{T} r_{t'}^k$ to enable autoregressive trajectory modeling during training. During inference, target RTG $G^\star$ is targeted as total return for an episode. 
This yields the following trajectory representation, which can be effectively used for autoregressive training and sequence generation:
\begin{equation}
\tau =
\left(
\hat{R}_0,\, s_0,\, a_0,\;
\hat{R}_1,\, s_1,\, a_1,\;
\ldots,\;
\hat{R}_T,\, s_T,\, a_T
\right)
\end{equation}
where $T$ denotes trajectory length. At timestep $t$,
DT takes a trajectory sequence $\tau$ autoregressively
as input, which contains the most recent K-step history (i.e., the number of
RTG–state–action tuples used to predict the next action). 
Please note that the objective of offline reinforcement learning is to learn a policy that maximizes the expected return
$\mathbb{E}\!\left[\sum_{t=0}^{T} r_t\right]$ in the underlying MDP.

In DT, the RTG, state, and action tokens are first mapped into a shared latent space and augmented with temporal (positional) embeddings $\pi(t)$ as:
\begin{equation}
u_{x,t} := \pi(t) \oplus \varphi_x(x_t), \quad x \in \{\hat{R}, s, a\} 
\end{equation}
where $\varphi_x(\cdot)$ denotes the modality-specific embedding function and $\oplus$ represents concatenation.
The resulting token embeddings are processed by stacked masked causal self-attention blocks, following the decoder-only generative pre-trained transformer (GPT) architecture \cite{NEURIPS2021_7f489f64}, which preserves the autoregressive structure by restricting attention to past and present tokens. As a result, decision making is performed through sequence generation, allowing policy learning to be framed as conditional next-token prediction over offline trajectories.
\begin{algorithm}[hbpt!]
\caption{Multi-Task Prompt-DT Training and Evaluation}
\label{alg:promptdt}
\begin{algorithmic}[1]
\REQUIRE Training $\mathcal{T}^{\text{train}}$, and evaluation tasks $\mathcal{T}^{\text{eval}}$, Prompt-DT $f_\theta$, iterations $N$, offline dataset $\mathcal{D}$, demonstrations $\mathcal{P}$, per-task batch size $M$, target returns $G^\star$, episode length $T$, prompt trajectory length $K\star$, trajectory length $K$ 
\STATE \textbf{Training}
\FOR{$n = 1$ to $N$}
    \FOR{each task $\mathcal{T}_i \in \mathcal{T}^{\text{train}}$}
        \FOR{$m = 1$ to $M$}
            \STATE Sample trajectory $\tau_{i,m} \sim \mathcal{D}_i$ 
            \STATE Sample prompt $\tau_{i,m}^\star = \text{GetPrompt} (\mathcal{T}_i, \mathcal{P}_i)$  
            \STATE Construct input $\tau_{i,m}^{\text{input}} = (\tau_{i,m}^\star, \tau_{i,m})$
        \ENDFOR
        \STATE Form minibatch $\mathcal{B}_i^M = \{\tau_{i,m}^{\text{input}}\}_{m=1}^M$
    \ENDFOR
    \STATE Form batch $\mathcal{B} = \bigcup \mathcal{B}_i^M$ for all $i$
    \STATE Predict actions $a^{\text{pred}} = f_\theta(\tau^{\text{input}}), \forall \tau^{\text{input}} \in \mathcal{B}$
    \STATE Compute masked Cross-Entropy loss $\mathcal{L_{CE}}$
    \STATE Update parameters $\theta \leftarrow \theta - \alpha \nabla_\theta \mathcal{L}$
\ENDFOR

\STATE \textbf{Evaluation}
\FOR{each task $\mathcal{T}_j \in \mathcal{T}^{\text{eval}}$}  
    \STATE Initialize history $\tau$ with zeros, desired reward $g = G^\star_j$
    \STATE Select prompt of length $K^\star$ from top-1 trajectory\\ 
    \hspace{1em}$\tau_j^\star = \text{GetBestPrompt}(T_j,\mathcal{P}_j)$    
    \FOR{$t = 1$ to $T$ } 
        \STATE Get action $a_t = f_\theta((\tau^\star, \tau))[-1]$
        \STATE Execute $a_t$ in env, get feedback $s_{t+1}, r_t$
        \STATE Update target reward $g \leftarrow g - r_t$
        \STATE Append $[s_{t+1}, a_t, g]$ to recent history $\tau$
    \ENDFOR
\ENDFOR
\end{algorithmic}
\end{algorithm}

In PromptDT (See Alg.\ref{alg:promptdt}), training and evaluation are performed using input sequences composed of a task-specific prompt trajectory $\tau^\star$ of length $K^\star$ with historical trajectory $\tau$ of length $K$, where $K > K^\star$ \cite{xu2022prompt} (lines 1–9). The constructed prompts are sampled stochastically from the trajectories to improve stability and reduce overfitting \cite{11080254}. It should be noted that minibatches are formed across all training tasks (lines 10–11), and PromptDT predicts actions in an autoregressive manner, conditioned on the combined sequence (line 12).
The model is trained by minimizing a masked cross-entropy loss over action tokens (e.g, based on number of UEs and BSs), followed by parameter updates via gradient descent (line 13-15).
Then, during the evaluation, a high-reward prompt is selected for each unseen task (lines 18–19) \cite{10827032}. Given a target return (e.g., task-specific max-reward in offline datasets), PromptDT generates actions in a closed-loop manner conditioned on the prompt and interaction history (lines 21–24), updating the RTG and trajectory until episode termination. An offline dataset $D$ is assumed to be available for each task during training, and demonstration trajectories $P$ are assumed to be accessible during both training and evaluation. Furthermore, the robustness of PromptDT to different prompt lengths has been demonstrated in~\cite{xu2022prompt}. In our implementation, we adopt the default configuration with historical trajectory length $K = 20$ and prompt length $K^\star = 5$.

\vspace{-5pt}
\section{Simulation Results}
\vspace{-5pt}
\subsection{Settings}

In our study, we construct a set of multi-task learning scenarios based on the open-source system-level lightweight simulation platform, namely mobile-env \cite{9789886}. We consider a total of 12 tasks, which we obtain by varying the scale of the network and scheduler policy used for resource allocation as shown in Tab. \ref{tab:task_scales}.
Due to the different task sizes, which lead to mismatched state and action space dimensionalities across tasks, we apply zero padding to all state and action representations so that they match the maximum dimensions.
Invalid action dimensions are then masked in the policy outputs during both training and inference. Also, ComP environment parameters are summarized in Tab. \ref{tab:promptdt_comp} following the default settings in \cite{9789886}. Experiments are conducted on the Kaggle platform, and NVIDIA Tesla P100 GPU is utilized.

\begin{table}[h!]
\centering
\caption{Multi-Task Configurations for 12 Tasks}
\label{tab:task_scales}
\begin{tabular}{lccccc}
\hline
\textbf{Scale} & \textbf{Scheduler(s)} & \textbf{\#UE} & \textbf{\#BS} & \textbf{State Dim.} & \textbf{Map Size} \\
\hline
Tiny       & ReF / PF & 5  & 3 & 35  & [200,200] \\
Small      & ReF / PF & 6  & 4 & 54  & [200,225] \\
SmallPlus  & ReF / PF & 7  & 4 & 63  & [200,225] \\
Medium     & ReF / PF & 10 & 5 & 110 & [200,250] \\
MediumPlus & ReF / PF & 12 & 6 & 156 & [200,275] \\
Large      & ReF / PF & 15 & 7 & 225 & [200,300] \\
\hline
\end{tabular}
\end{table}
\subsubsection{Baselines}
We compare PromptDT with PPO-based baselines. PromptDT models are instantiated with approximately 5M, 15M, and 45M parameters by varying the hidden dimension. In addition, models are trained under different multi-task settings using 8, 10, and 12 tasks. The training trajectories are collected from PPO agents, with default settings, and include both suboptimal and successful interactions with the environment. Using only successful interactions makes it harder for the PromptDT model to distinguish desirable trajectories from suboptimal ones, since the absence of failure cases removes important contrastive signals during training. During the evaluation, the single-task expert PPO agents are compared against the multi-task PromptDT models. Overall, this results in nine PromptDT variants spanning different training task counts and model capacities.

\begin{table}[h]
\centering
\caption{PromptDT and CoMP environment parameters.}
\label{tab:promptdt_comp}
\setlength{\tabcolsep}{4pt}
\renewcommand{\arraystretch}{1.0}
\begin{tabular}{@{}l c p{2.5cm} c@{}}
\toprule
\textbf{PromptDT params.} & \textbf{Value} & \textbf{CoMP params.} & \textbf{Value} \\
\midrule
n\_heads        & 4              & $f_c$ (GHz)              & 2.5 \\
n\_blocks       & 6              & BW (MHz)                 & 9 \\
$K$, $K^\star$  & 20, 5           & $P_{Tx}$ (dBm)           & 30 \\
batch\_size     & 16             & $N_0$ (dBm/Hz)           & $-90$ \\
optimizer       & AdamW          & $\rho_{min}$ (dB)        & $-77$ \\
lr\_rate        & $10^{-4}$      & Cell tower height (m)    & 50 \\
weight\_decay   & $10^{-4}$      & $(w_1,w_2,w_3)$          & $(10,0,10)$ \\
hidden\_dim(s)  & 256,448,792    & $(U_{\min}, U_{\max})$   & $(-20, 20)$ \\
training steps  & 50\_000        & UE Velocity (m/s)        & 1.5 \\
\bottomrule
\end{tabular}
\end{table}

\subsection{Results}
\begin{figure}[h!]
    \centering
    \includegraphics[width=0.85\columnwidth, height=6.2cm]{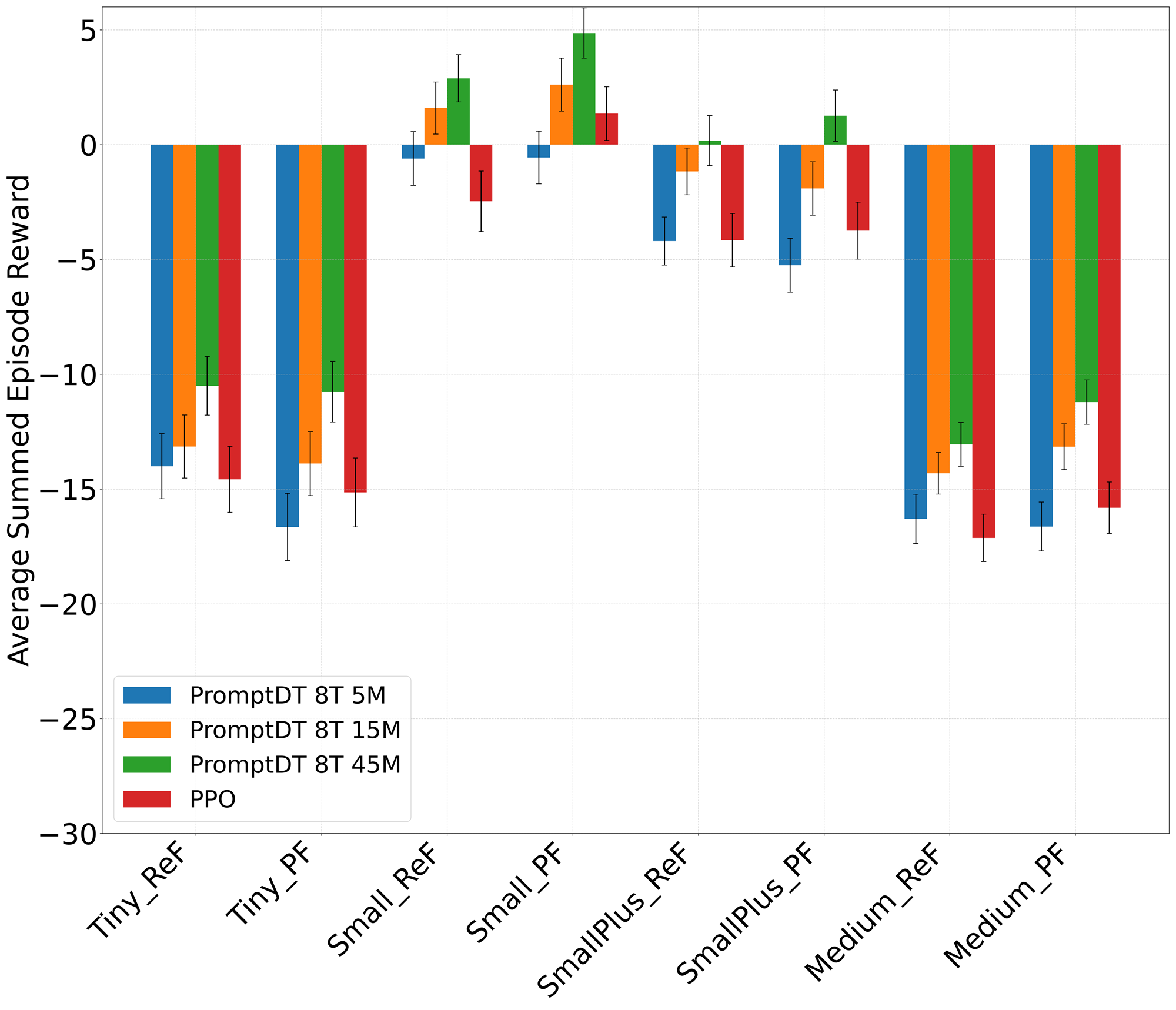}
    \caption{PromptDT trained and tested on 8 tasks.}
    \label{fig:reward_8} 
    \vspace{-15pt}
\end{figure}

\begin{figure}[h!]
    \centering
    \includegraphics[width=0.85\columnwidth, height=6.2cm]{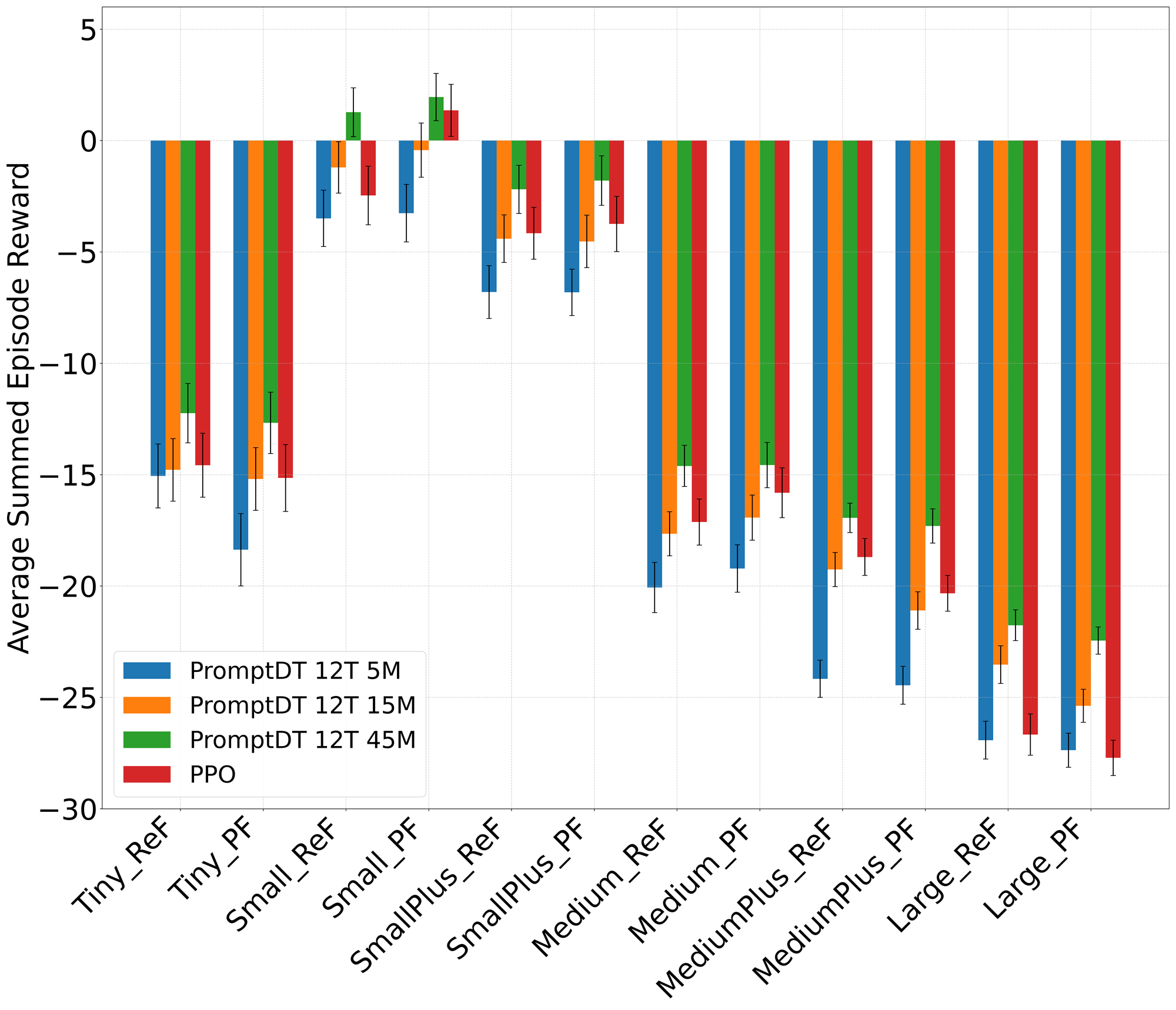}
    \caption{PromptDT trained and tested on 12 tasks.}
    \label{fig:reward_12} 
\end{figure}

Figures ~\ref{fig:reward_8} and \ref{fig:reward_12} compares the average QoE-based episode reward of PromptDT and PPO across all task scales in Table~\ref{tab:task_scales}. Results report the mean and standard error of the mean over 100 test episodes with different seeds. PPO is trained per task, while PromptDT is jointly trained over $T$ tasks and evaluated individually. Absolute reward levels vary across tasks due to differing contention and allocation constraints, with harder environments naturally yielding lower rewards. This mainly stems from UE/BS ratio, BS layouts and dimension of state and action space (i.e., harder decision-making).

Across all configurations, PromptDT consistently outperforms single-task PPO once model capacity exceeds 15 M parameters. In small-scale environments (e.g., Tiny, Small, SmallPlus), gains are modest but stable (5–10\%), 
As environment scale increases, the performance gap widens. In Medium and MediumPlus settings, PromptDT improves rewards by approximately 12–20\%, benefiting from better generalization in denser and higher-dimensional scenarios. The largest gains occur in the Large-scale environment, where PromptDT outperforms PPO by up to 25–30\% under severe contention and stringent allocation constraints, as its sequence-modeling formulation enables more effective learning in large, high-dimensional state and action spaces, where single-task PPO struggles to generalize. PromptDT also adapts effectively across different allocation strategies (e.g., ReF and PF), maintaining consistent performance gains despite their distinct scheduling behaviors. Increasing the number of training tasks makes joint generalization more challenging due to gradient conflicts and can slightly reduce per-task performance under a massive number of tasks \cite{kong2025mastering}; nevertheless, PromptDT trained on 12 tasks still surpasses PPO in most environments.

\begin{figure}[h!]
    \centering
    \includegraphics[width=0.55\columnwidth]{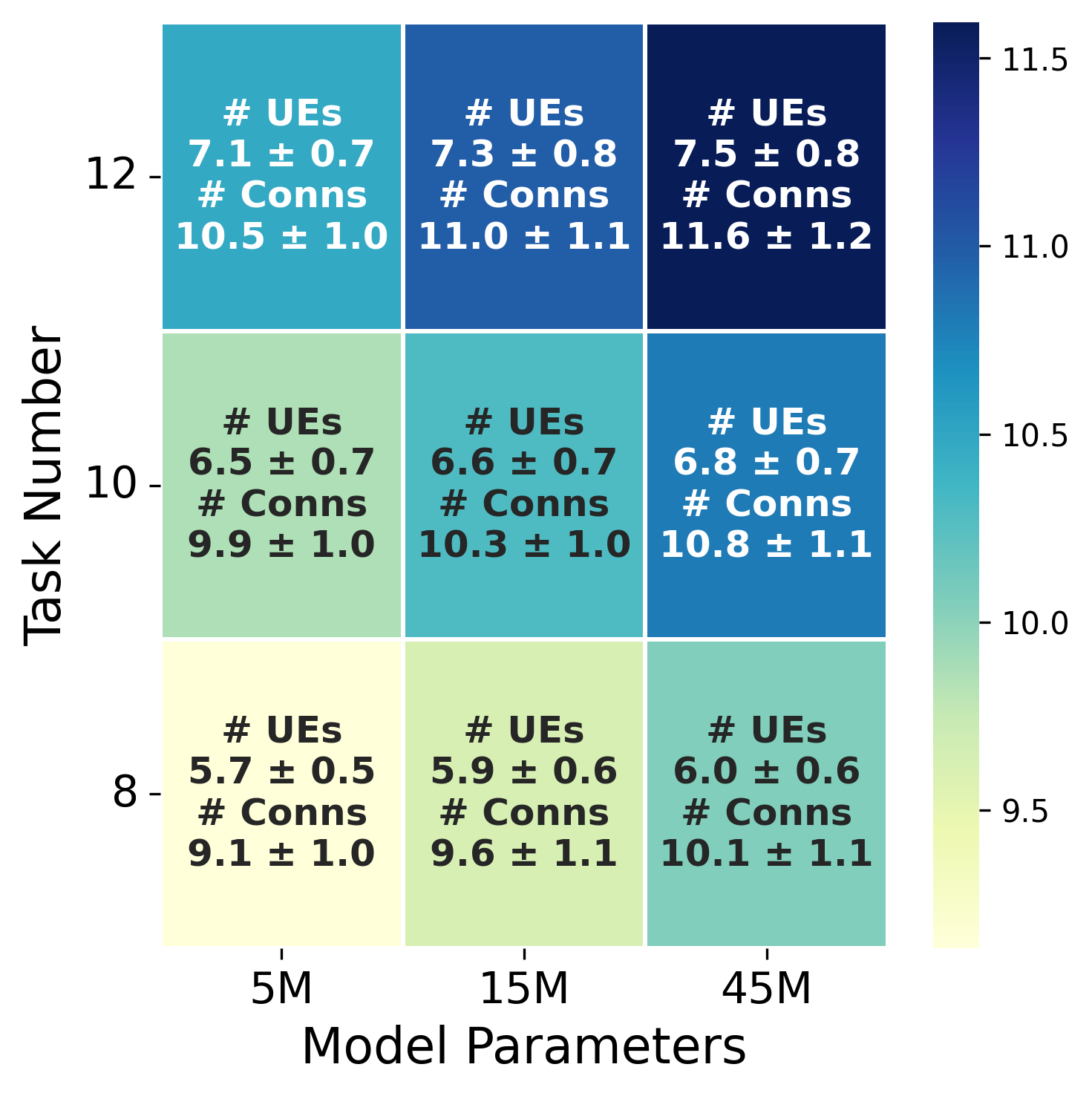}
    \caption{Avg. Number of Active UEs and Established Links}
    \label{fig:connections}
\end{figure}

As illustrated by Figures \ref{fig:reward_8}, \ref{fig:reward_12}, and \ref{fig:connections}, increasing the model capacity improves the overall task score, even as the number of tasks increases. Larger models achieve higher average episode rewards across all settings, including the 12-task case, indicating improved robustness under multi-task complexity.
Fig.~\ref{fig:connections} reports the number of active UEs and the number of established links between UEs and BSs. Configurations with 12 tasks naturally exhibit more connections due to larger environments involved, yet higher-capacity models (e.g., 15M and 45M models) still support more connected UEs and links compared to the models with 5M parameters, highlighting the benefit of model scaling under higher contention in CoMP.\\

\begin{figure}[h!]
    \centering
    \includegraphics[width=1\columnwidth]{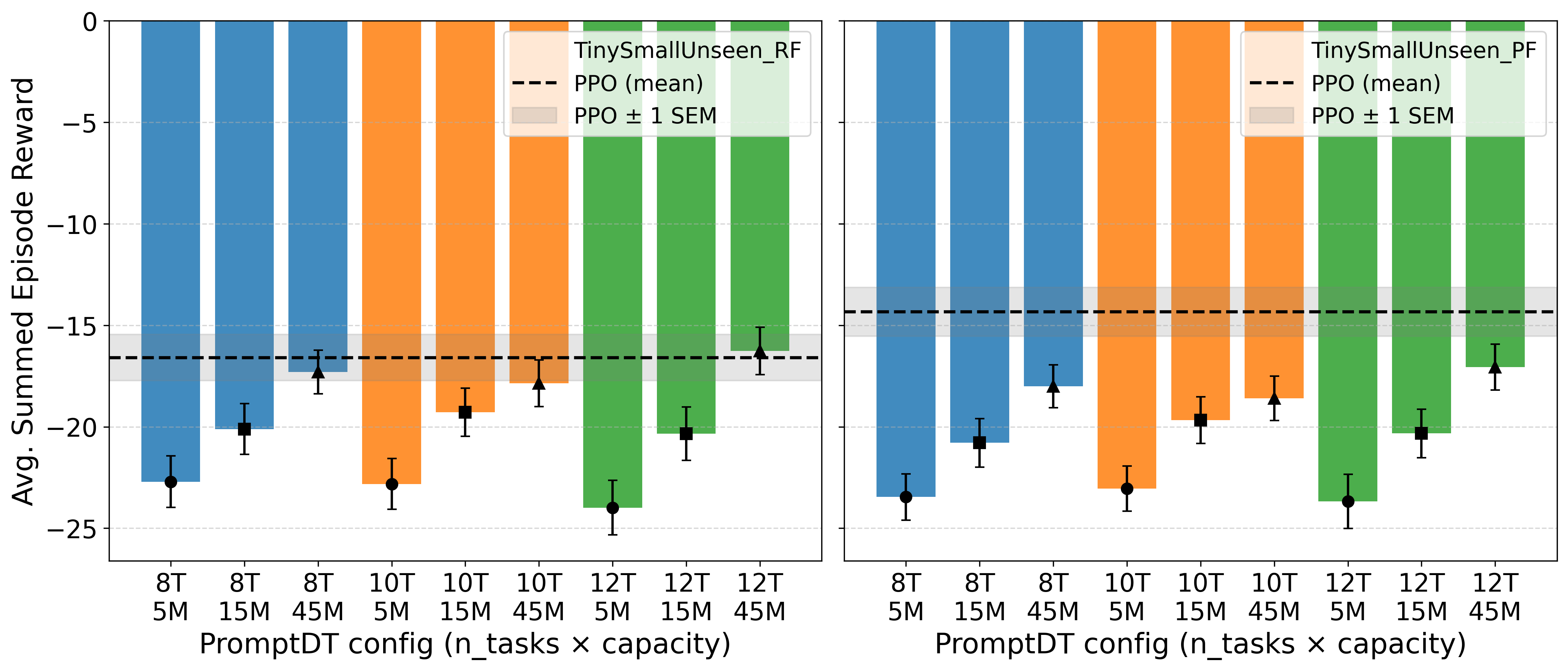}
    \caption{Unseen Task Performance
    }
    \label{fig:reward}
\end{figure} 
\textbf{Unseen Task Performance:} In Fig. \ref{fig:reward}, we report average reward on unseen tasks as a function of model capacity of PromptDT and the dashed line indicates the performance of PPO trained directly on the target environments. PromptDT, evaluated on unseen tasks, approaches or even exceeds this trained PPO baseline as model capacity increases, demonstrating effective few-shot adaptation without task-specific retraining.

Beyond performance gains, PromptDT also offers advantages in deployment. The centralized PPO-based solution, DeepCoMP, in \cite{schneider2023deepcomp} is proposed to be deployed by partitioning the entire network into multiple areas (e.g., based on cell coverage or operational criteria), each controlled independently by a dedicated DeepCoMP agent. In contrast, PromptDT adopts a multi-task modeling paradigm, enabling a single unified model to generalize across multiple areas and task configurations. Instead of training and maintaining separate agents for each area, which increases deployment and maintenance complexity, PromptDT leverages task-specific prompts to adapt to different operational scenarios or areas.

\begin{table}[h!]
\centering
\caption{Complexity comparison of PromptDT models.}
\label{tab:complexity}
\setlength{\tabcolsep}{3pt}
\footnotesize
\begin{tabular}{lcccc}
\toprule
\textbf{Model} &
\shortstack{\textbf{FP16}\\\textbf{Mem.}} &
\shortstack{\textbf{Inf.}\\\textbf{Time (ms)}} &
\shortstack{\textbf{GFLOPs}\\\textbf{/Inf.}} &
\shortstack{\textbf{GFLOPs}\\\textbf{/Token}} \\
\midrule
Prompt-DT 5M  & 10 MB & 10.68 & 0.749 & 0.010 \\
Prompt-DT 15M & 30 MB & 11.31 & 2.240 & 0.030 \\
Prompt-DT 45M & 90 MB & 11.01 & 6.902 & 0.092 \\
\bottomrule
\end{tabular}
\end{table}

\textbf{Complexity Analysis:} The proposed PromptDT models achieve computational efficiency for multi-cell selection in CoMP systems. As shown in \ref{tab:complexity}, they require at most 90 MB FP16 memory, with computational costs ranging from 0.749 to 6.902 GFLOPs per inference at full context length (e.g., 75 tokens including states, actions and RTGs), while maintaining an inference latency of approximately 11 ms on low-end GPUs, demonstrating practical feasibility for real-time RRM deployment.

\vspace{-5pt}
\section{CONCLUSION}
This paper introduced a PromptDT-based multi-task learning framework in CoMP domain to overcome the inherent limitations of conventional RL, such as training instability, poor generalization, and the prohibitive costs of retraining under varying network configurations. Our results demonstrate that this architecture outperforms single-task PPO baselines by up to 49\% in QoE, while maintaining seamless scalability through model capacity expansion. By leveraging task-specific prompts, the framework achieves robust few-shot adaptation to unseen tasks. 
Future work may explore techniques such as mixture-of-experts to mitigate gradient-conflict across tasks to further increase generalization and task-scalability. 
\vspace{-5pt}
\section*{ACKNOWLEDGEMENT}
This work has been supported by the NSERC Canada Research Chairs Program.

\renewcommand\refname{References}
\addcontentsline{toc}{section}{References}
\bibliographystyle{IEEEtran}
\bibliography{ref}

\end{document}